\documentclass{iau}
\usepackage{graphicx}
\usepackage{caption}

\title[Molecules in the CND] 
{ Molecules in the Circumnuclear Disk of the Galactic Center}

\author[Harada et al.]   
{Nanase Harada$^1$, Denise Riquelme$^1$, 
 Serena Viti$^2$,
 Karl Menten$^1$, Miguel Requena-Torres$^1$, Rolf G\"usten$^1$, Stefan Hochg\"urtel$^1$
 }

\affiliation{$^1$Max Planck Institute for Radio Astronomy, Auf dem H\"{u}gel 69, D-53121, Bonn \\ email: {\tt harada@mpifr-bonn.mpg.de} \\[\affilskip]
$^2$University College London, Department of Physics and Astronomy, Gower Street, London WC1E 6BT, UK}

\pubyear{2013}
\volume{303}  
\pagerange{}
\jname{The Galactic Center: Feeding and Feedback in a Normal Galactic Nucleus}
\editors{L. Sjouwerman, J. Ott \& C. Lang, eds.}
\begin{document}

\maketitle

\begin{abstract}
Within a few parsecs 
around  the central Black Hole Sgr A*, chemistry in the dense molecular cloud material of the circumnuclear disk (CND) can be affected by many 
energetic phenomena such as high UV-flux from the massive central star cluster, X-rays from Sgr A*, shock waves, and an
enhanced cosmic-ray flux.
 Recently, spectroscopic surveys with the IRAM 30 meter and the APEX 12 meter telescopes of substantial parts of the 80--500 GHz frequency range were made toward selected positions  in and near the CND
 These datasets contain lines from the molecules 
HCN, HCO$^+$, HNC, CS, SO, SiO, CN, H$_2$CO, HC$_3$N, N$_2$H$^+$, H$_3$O$^+$ and others. We conduct Large Velocity Gradient analyses to obtain column densities  and total hydrogen densities, $n$, for each species in molecular clouds located in the southwest lobe of CND. 
The data for the above mentioned molecules indicate 10$^5\,$cm$^{-3} \lesssim n <10^6\,$cm$^{-3}$, which shows that the CND is tidally unstable.
The derived chemical composition is compared with a chemical model calculated using the UCL\_CHEM code that includes gas and grain reactions, and the effects of shock waves. Models are run for varying shock velocities, cosmic-ray ionization rates, and number densities. The resulting chemical composition is fitted best to an extremely high value of cosmic-ray ionization rate  $\zeta \sim 10^{-14}\,$s$^{-1}$, 3 orders of magnitude higher than the value in regular Galactic molecular clouds, if the pre-shock density is $n=10^5\,$cm$^{-3}$. 
\end{abstract}
\firstsection 
\section{Introduction}
The central molecular zone (CMZ) in the central few hundreds parsecs in the Galactic Center hosts a large amount of molecular gas ($M\gtrsim 2 \times 10^7\,M_{\odot}$; \cite[Oka et al. 1998]{1998ApJ...493..730O}). The interaction between the molecular mass and high-energy processes in the Galactic Center region has been observed in many ways. $Fermi$-LAT captured high-energy gamma-ray emission from the central regions of the Galaxy \cite[(Chernyakova et al. 2011)]{2011ApJ...726...60C}, which suggest the presence of an enhanced cosmic-ray density. Observation of X-rays suggest that the supermassive black hole was active 100-300 years ago \cite[(Ponti et al. 2010]{2010ApJ...714..732P} and \cite[Koyama et al. 1996)]{1996PASJ...48..249K}. Chemical compositions of molecular clouds in the CMZ is likely to be affected by these high-energy activities.
Sgr A* is surrounded by the circumnuclear disk (CND), a ring of molecular/atomic gas at a distance of 2-7 pc from the location of the supermassive black hole, whose chemistry can be most affected by the activity of Sgr A*. Molecules with relatively strong emissions such as CO, HCN, HCO$^{+}$ have been previously observed both by single-dish and interferometers (e.g., \cite[Serabyn et al. 1986]{1986A&A...169...85S},  \cite[Christopher et al. 2005]{2005ApJ...622..346C}). To study the overall chemical composition of the CND, a spectral line survey has been conducted that covers most of the frequency range between 80 and 500 GHz that is observable from the ground (PI: Denise Riquelme). Here we outline how the observed data were analyzed in a framework of a chemical model to constrain the physical conditions in the CND.

\section{Physical Condition of the Source}
Density estimates for the molecular clouds in the CND found in the literature vary greatly. \cite{2005ApJ...622..346C} derived an average density of $n$, of $(3-4)\times 10^7\,$cm$^{-3}$ from their HCN(1-0) emission map. Estimates by \cite{2009ApJ...695.1477M} also show densities of $n > 10^7\,$cm$^{-3}$. On the other hand, a much lower density of $n \sim 10^{4} - 10^{5}\,$cm$^{-3}$ was derived from multi-J CO line observation by \cite{2012A&A...542L..21R}. Relatively high temperatures of $T > 200\,$K have been reported by  \cite{2012A&A...542L..21R} and \cite{2005ApJ...623..866B}. 

 In the CND, sources of ionization may be cosmic-rays, X-rays, and UV-photons. Ionized atoms or molecules can induce ion-neutral reactions, which in general are faster than neutral-neutral reactions. UV-photons, which either come from OB stars or internally generated by cosmic-ray or X-rays, can efficiently dissociate molecules.

The value of the cosmic-ray ionization rate in most parts of the Galaxy is thought to be $\zeta \sim 3 \times 10^{-17}\,$s$^{-1}$ in dense molecular clouds and about an order of magnitude higher for diffuse clouds \cite[(Indriolo \& McCall 2012)]{2012ApJ...745...91I}. For CMZ, there are several claims of an elevated cosmic-ray ionization rate. In the dense cloud of the Sgr B2 region, \cite{2006A&A...454L..99V} showed that the cosmic-ray ionization rate is $\zeta \sim 4 \times 10^{-16}\,$s$^{-1}$. \cite{2008ApJ...688..306G} found that the value in more diffuse CMZ material is $\zeta > 10^{-15}\,$s$^{-1}$. There are even higher claims, such as $\zeta \sim 5 \times 10^{-13}\,$s$^{-1}$ \cite[(Yusef-Zadeh et al. 2007)]{2007ApJ...656..847Y}.

X-rays affect chemistry in a similar way as cosmic-rays by a primary or secondary ionization followed by internal generation of UV-photons, which dissociate molecules. The X-ray luminosity of the supermassive black hole in Sgr A* currently has an X-ray luminosity, $L_X$, of $\sim 2 \times 10^{33}\,$erg s$^{-1}$ in the 2-10 keV range with the energy dependence $n(E) \propto E^{\Gamma}$ having a photon index $\Gamma$ of 2.7 \cite[(Baganoff et al. 2003)]{2003ApJ...591..891B}. We estimated a value for the X-ray ionization rate using the method described in \cite{1996ApJ...466..561M}. The density of the cloud is assumed to be $n=10^5\,$cm$^{-3}$, and the X-ray flux at $E_X < 2\,$keV is assumed to have an energy dependence as exp(-2keV/E) and the minimum energy is taken as 0.1 keV. The X-ray ionization rate is extremely high ($\zeta \sim 10^{-14}\,$s$^{-1}$) for the very edge of the cloud, but rapidly falls below the value of the cosmic-ray ionization rate in regular dense clouds in the Galaxy, which means that presently X-rays cannot be the dominant driving force of the chemistry although, the possibly, a higher activity in the past may have had a strong impact. 

 Star clusters within the central parsec from Sgr A* are emitting ionizing photons at rate of $2\times 10^{50}$s$^{-1}$\cite[(Lacy et al. 1980)]{1980ApJ...241..132L}, producing photon-dominated regions (PDRs). Although a significantly enhanced photon field is expected, UV-photons can be attenuated with column densities of $N \sim 10^{22}\,$cm$^{-2}$. Since UV-photons may be attenuated deep in the molecular cloud, the effect of UV-photons are treated separately in our model.


\section{Observation and Analysis}
Data was taken with IRAM 30 meter telescope and the FLASH receiver on the APEX telescope (for details, Riquelme et al., in preparation). Overlaid on the interferometer data of HCN(4-3) by \cite{2009ApJ...695.1477M} in color scale, Figure \ref{fig:beam} shows the position the southwest lobe of the CND towards which our line survey data were taken indicated by solid circles for the largest and smallest beam size of IRAM 30 m, and dotted circles for APEX FLASH receiver. The beam temperature is converted to a main brightness temperature scale assuming source sizes of 0.3 pc from the interferometer data of \cite{2009ApJ...695.1477M}. In the same line of sight, there are multiple velocity components (50 km s$^{-1}$ cloud, 20 km s$^{-1}$ cloud, and CND), and only a negative velocity component, which certainly comes from the CND is used for our analysis. Column densities for each species are derived employing a large velocity gradient (LVG) analysis based on non-LTE radiative transfer code Radex \cite[(van der Tak et al. 2007)]{2007A&A...468..627V}. Column densities are obtained for HCN, HCO$^+$, HNC, CS, SO, SiO, CN, H$_2$CO, HC$_3$N, N$_2$H$^+$, H$_3$O$^+$. The best-fit densities obtained by the LVG analysis for these species are $10^5 \lesssim n <10^6\,$cm$^{-3}$.

\section{Chemical Models}
\begin{figure}[h]

\begin{minipage}[c]{.5\textwidth}
  \centering
 
 \includegraphics[width=.7\linewidth]{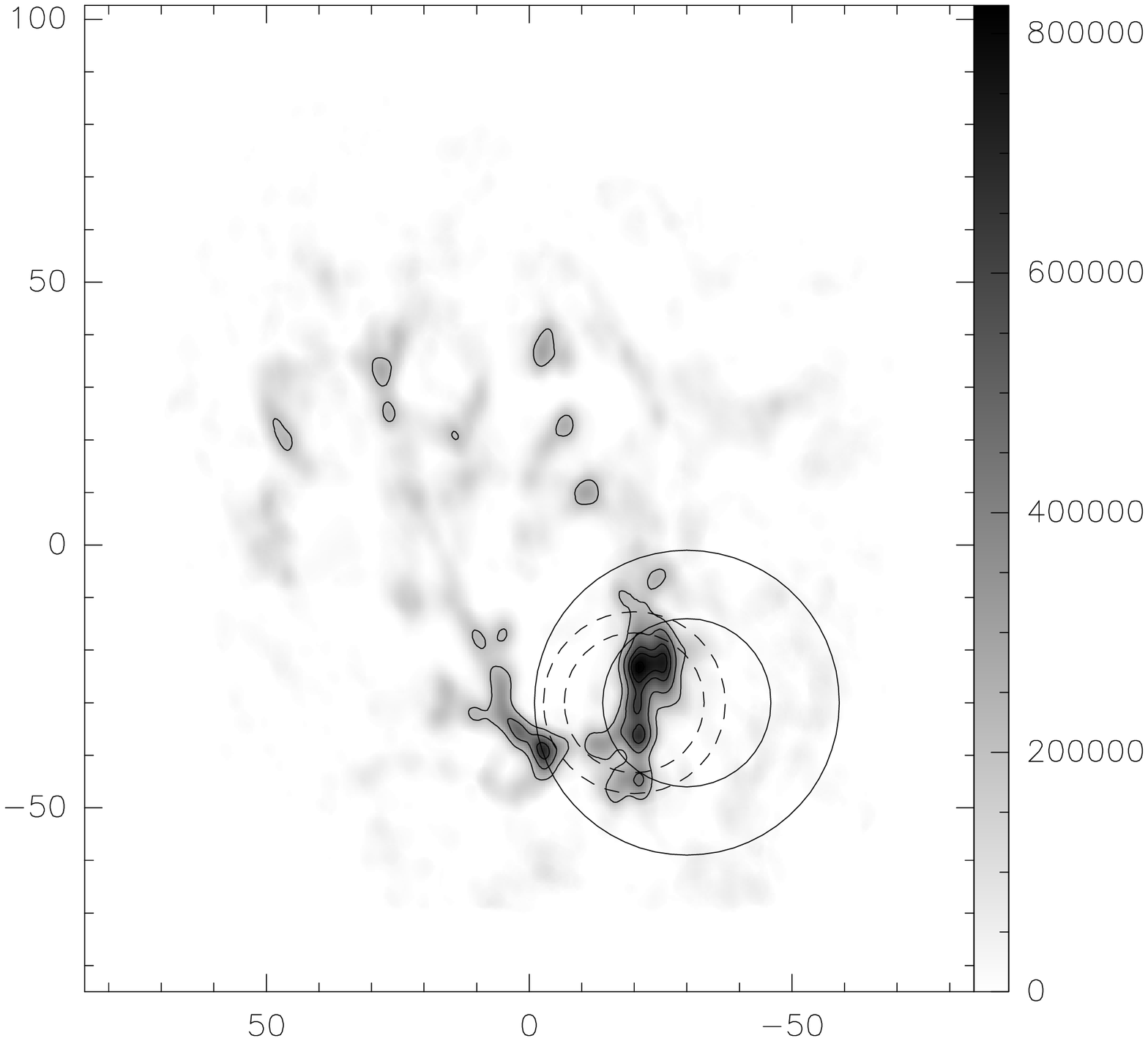} 
  \captionof{figure}{Grey scale shows the intensity of HCN(4-3) emission from \cite{2009ApJ...695.1477M}. Beam sizes of IRAM and APEX telescopes are shown in solid circles and dotted circles, respectively.}
  \label{fig:beam}
\end{minipage}~~~~%
\begin{minipage}[c]{.5\textwidth}
  \centering
  \includegraphics[width=\linewidth]{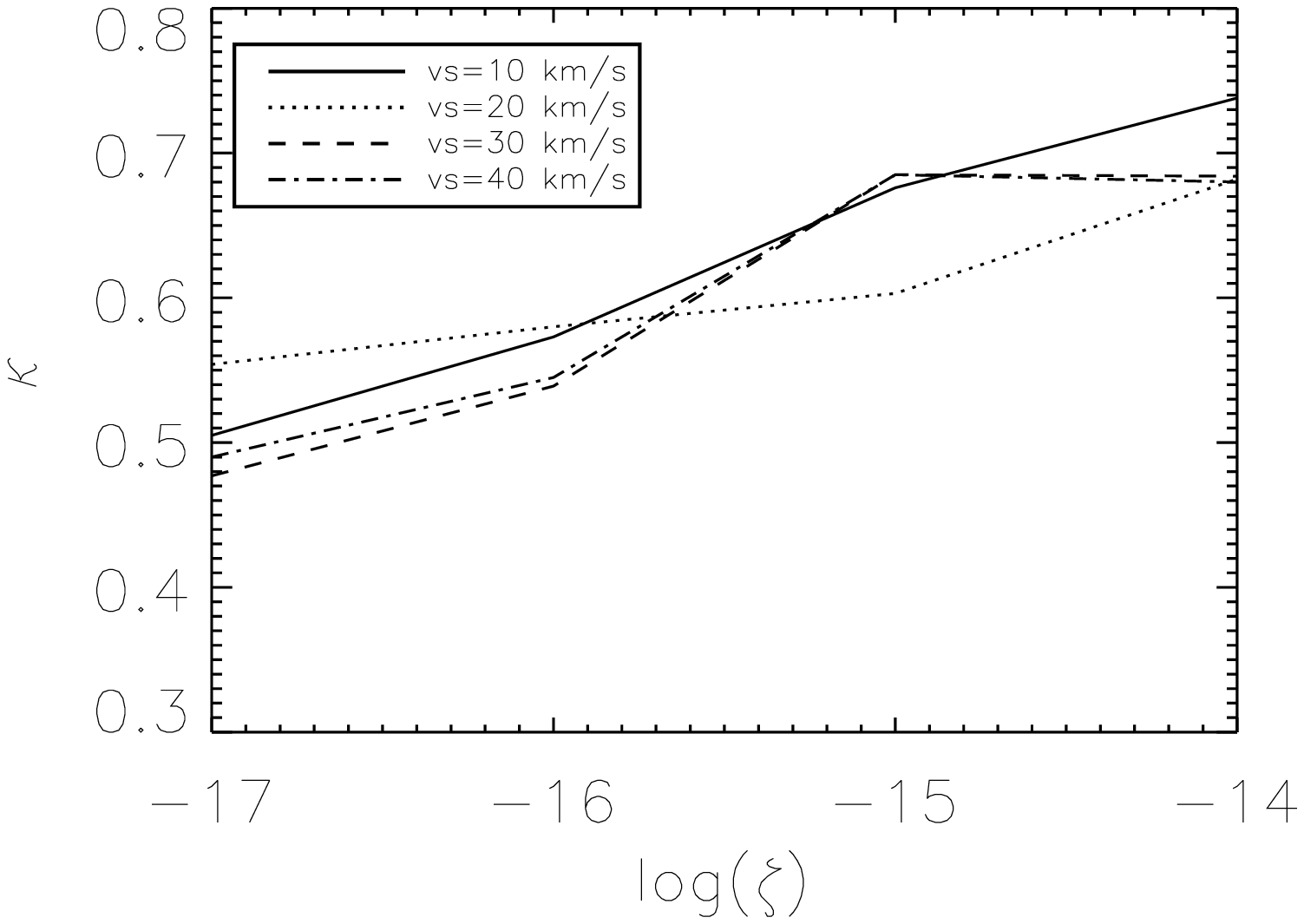}
  \captionof{figure}{Values of confidence level with varying shock velocities are shown for the density $n=10^5\,$cm$^{-3}$.}
  \label{fig:fit}
\end{minipage}

\end{figure}

We used time-dependent gas-grain chemical model UCL\_CHEM \cite{2004MNRAS.354.1141V}, which was extended to include the effect of shock waves \cite{2011ApJ...740L...3V}. The model calculations were run for different final densities $n=10^4,~10^5$, and $10^6\,$cm$^{-3}$, cosmic-ray ionization rate $\zeta = 10^{-17}$, $10^{-16}$, $10^{-15}$, and $10^{-14}\,$s$^{-1}$, and shock velocities of 10, 20, 30, and 40 km s$^{-1}$. Different sets of elemental abundances were used, including the "low-metal" abundances and solar abundances described in \cite{2008ApJ...680..371W}. The overall match with the observations is examined by a parameter $\kappa$ used in \cite{2007A&A...467.1103G};
$\kappa_{\rm i}=erfc \left[|X_{i, calc}-X_{\rm i, obs}|/(\sqrt{2} \sigma) \right].$
Figure \ref{fig:fit} shows values of $\kappa$ in each model at its time of best agreement. The overall agreement with the observation is better with high value of cosmic-ray ionization rate $\zeta = 1 \times 10^{-14}\,$s$^{-1}$ for all the densities, but the $\zeta$-dependence is weaker when $n=10^4\,$cm$^{-3}$.  The best agreement in chemistry is achieved when $n=10^4\,$cm$^{-3}$, which is lower than the density derived by the LVG analysis.

\begin{figure}[h]
\begin{center}
\centerline{
\includegraphics[width=0.4\textwidth]{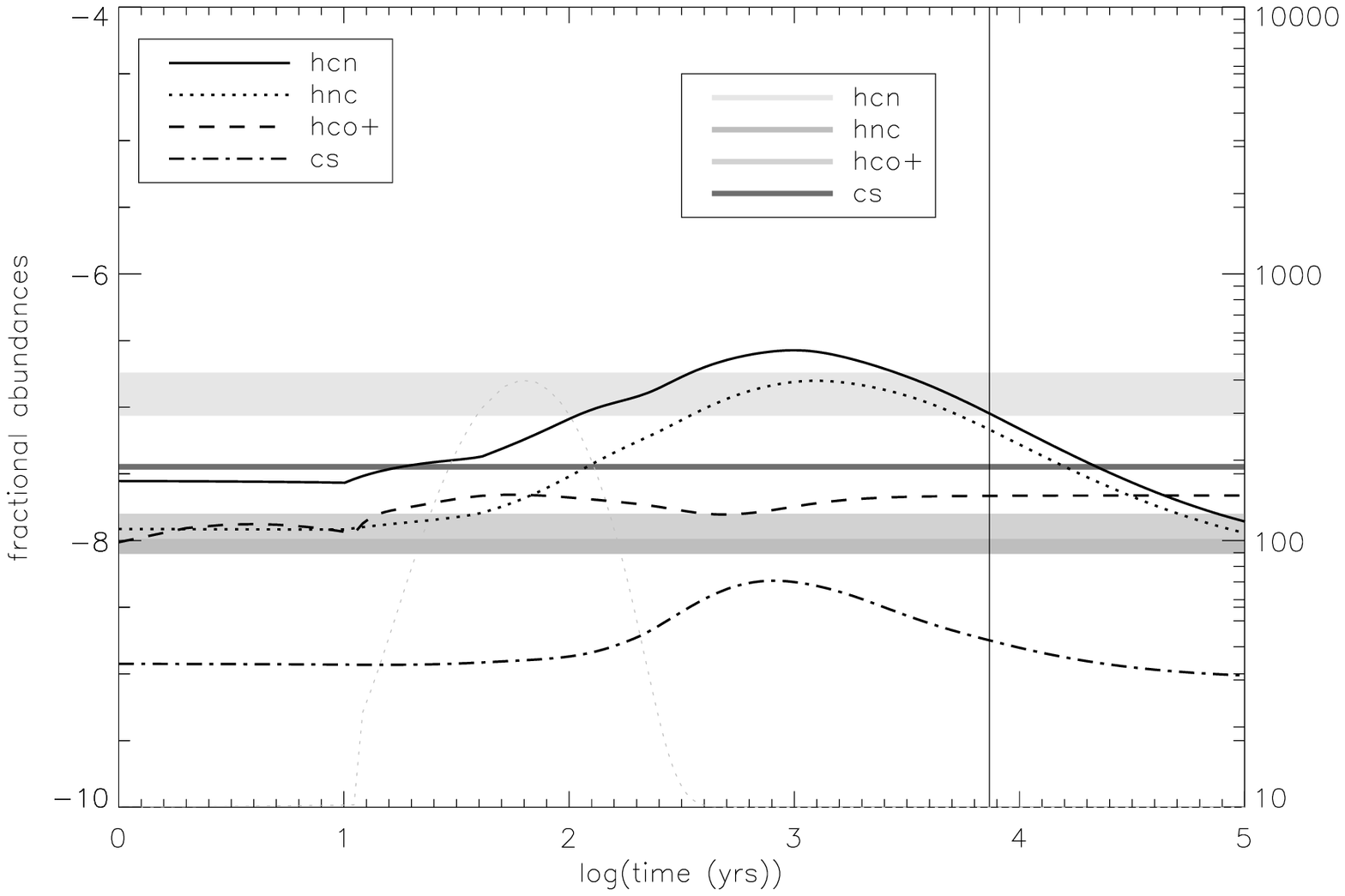}
\includegraphics[width=0.4\textwidth]{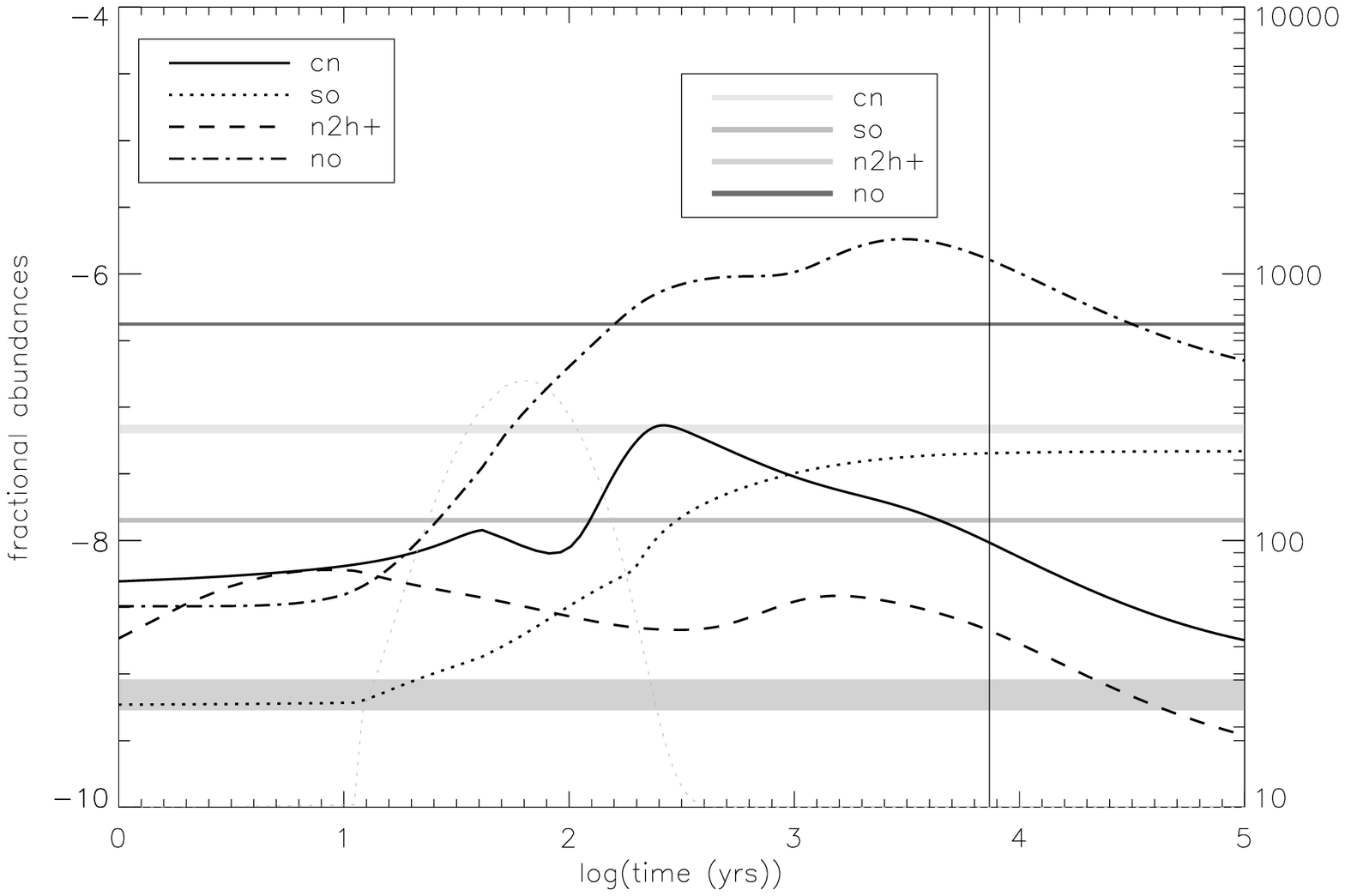}}
 \caption{Calculated fractional abundances averaged over time from the time of shock passage and observed values are shown for $n=10^5\,$cm$^{-3}$, $\zeta =10^{-14}\,$s$^{-1}$, and $v_s = 20\,$km s$^{-1}$.}
 \label{fig:abun}
\end{center}
\end{figure}

Figure \ref{fig:abun} shows fractional abundances of selected species as a function of time from the time of shock passage. Although for most species, the values are in good agreement with the observation at the time of best agreement at $t=10^4\,$yr, CN and CS are underabundant by more than an order of magnitude. The fractional abundance of CN can be enhanced in PDRs. There is also an observational claim that sulfur elemental abundances are higher in PDRs because of less depletion of sulfur on dust grains \cite{2006A&A...456..565G}.

\section{Discussion and Summary}
Our results show that, at the time of best agreement ($10^4\,$yr), the observed chemical composition fits chemical models best when the cosmic-ray ionization rate at the circumnuclear disk is $\zeta = 1 \times 10^{-14}\,$s$^{-1}$, about three orders of magnitude higher than the regular galactic molecular clouds. Although previous work has suggested the value similar to our result, the latter should be taken with caution. Our parsec-scale size beams cover multiple molecular cloud components that are resolved in interferometer maps, and that have varying densities, while our chemical models are run only for a fixed density in each model. The time of best agreement we derived happens to correspond to the time of cosmic-ray turn on suggested by the model of \cite{2011ApJ...726...60C}.



\end{document}